\documentclass[twocolumn,superscriptaddress,showpacs,prl]{revtex4}
\usepackage{graphicx}
\usepackage[english]{babel}

\usepackage[usenames]{color}

\begin{document}

\title{Damage in graphene due to electronic excitation induced by highly charged ions\\}

%{\small Version 15.10.2013  }}

%\author{}
\author{J.~Hopster}
\affiliation{Fakult\"at f\"ur Physik and CeNIDE, Universit\"at Duisburg-Essen, 47048 Duisburg, Germany}
\author{R.~Kozubek}
\affiliation{Fakult\"at f\"ur Physik and CeNIDE, Universit\"at Duisburg-Essen, 47048 Duisburg, Germany}
\author{B.~Ban-d'Etat}
\affiliation{CIMAP (CEA-CNRS-ENSICAEN-UCBN), 14070 Caen Cedex 5, France} 
\author{S.~Guillous}
\affiliation{CIMAP (CEA-CNRS-ENSICAEN-UCBN), 14070 Caen Cedex 5, France}
\author{H.~Lebius}
\affiliation{CIMAP (CEA-CNRS-ENSICAEN-UCBN), 14070 Caen Cedex 5, France}
\author{M.~Schleberger}
\email{marika.schleberger@uni-due.de}
\affiliation{Fakult\"at f\"ur Physik and CeNIDE, Universit\"at Duisburg-Essen, 47048 Duisburg, Germany}

\begin{abstract}
Graphene is expected to be rather insensitive to ionizing particle radiation. We demonstrate that single layers of exfoliated graphene sustain significant damage from irradiation with slow highly charged ions. We have investigated the ion induced changes of graphene after irradiation with highly charged ions of different charge states ($q$ = 28-42) and kinetic energies ($E_{\rm kin}$ = 150-450~keV). Atomic force microscopy images reveal that the ion induced defects are not topographic in nature but are related to a significant change in friction. To create these defects, a minimum charge state is needed. In addition to this threshold behaviour, the required minimum charge state as well as the defect diameter show a strong dependency on the kinetic energy of the projectiles. From the linear dependency of the defect diameter on the projectile velocity we infer that electronic excitations triggered by the incoming ion in the above-surface phase play a dominant role for this unexpected defect creation in graphene.  
\end{abstract}

\maketitle

%\section{Introduction}
The bombardment of surfaces with highly charged ions (HCI) offers an elegant way to deposit a large amount of energy into a very small volume and to study the behaviour of solid matter under such extreme conditions. In contrast to singly charged ions, HCI store a considerable amount of potential energy $E_{\rm pot}$ (defined as the energy to remove the electrons from the initial atom to infinity). Upon impact they trigger significant electronic excitations in the near surface region, which can result in a variety of nanoscaled structural modifications~\cite{Aumayr.2011}. For two dielectric materials, CaF$_2$ and KBr, the dependency of the structural surface modification on the charge state $q$ and the kinetic energy $E_{\rm kin}$ of the projectile was investigated in great detail~\cite{Heller.2008,ElSaid.2008,ElSaid.2012b} and could be explained in terms of different phase transitions and defect agglomeration mechanisms, respectively. In the case of CaF$_2$, the good agreement with model calculations based on a two-temperature-model~\cite{ElSaid.2008} suggests, that the transfer of the primary electronic excitation to the lattice by electron-phonon-coupling is the origin of the structural modifications. For conductive materials, however, the situation is less clear as experimental evidence is sparse~\cite{Tona.2007c,Pomeroy.2007c} and inconclusive. One way to investigate the fundamental interaction mechanisms of HCI with conductive targets is to use crystalline graphite. It can be used either in its bulk form (highly oriented pyrolytic graphite, HOPG) or in its two-dimensional (2D) form, graphene. Chemically identical, graphene shows superior electronic and heat transfer properties. As a consequence, it has even been proposed that graphene should be virtually transparent to energetic ions~\cite{Lehtinen.2010}. 

In this paper we show, that graphene sustains significant damage even by individual HCI impacts. We demonstrate that HCI irradiation can be used to create regions of enhanced friction, which we attribute to chemical changes of the graphene lattice and whose size can be easily controlled. We present evidence that this damage is due to the electronic excitation triggered by the projectile and that, in contrast to graphite, it only occurs above a certain potential energy, i.e.~above a threshold $E_{\rm pth}$. Furthermore, it turns out that the onset of defect formation in graphene as well as the defect size strongly depend on the ion's kinetic energy. These findings suggest an entirely different defect creation mechanism as in the case of formerly investigated bulk materials.

Highly charged ion irradiations were performed at the Duisburg beamline HICS~\cite{Peters.2009}. In this experiment, Xe$^{q+}$ ions with charge states $q$=28, 30, 32, 35, 38, 40, and 42 (covering a range of $E_{\rm pot}$ from 12 to 45~keV) were used with the kinetic energy kept constant at $E_{\rm kin}=260$~keV. With fixed charge states 28 and 30, respectively, the kinetic energy was varied from 150~keV to 450~keV, which was partially performed at ARIBE, located at GANIL in Caen. The irradiation took place under perpendicular incidence and typical fluences of $1 \times 10^9$~ions/cm$^2$ were used resulting in non-overlapping impacts. High quality graphene samples were exfoliated from HOPG on 90~nm amorphous SiO$_2$ on Si, yielding single layer graphene (SLG) and thick few-layer graphene (FLG) flakes on the same sample enabling direct comparison. Characterization before irradiation included Raman spectroscopy for SLG verification and atomic force microscopy (AFM) operated in contact mode for pre-inspection. The unique shape of indiviual flakes enabled us to compare the exactly same region before and after irradiation. For detailed defect analysis the AFM was operated in friction force mode (FFM)~\cite{Grafstrom.1994}, which is sensitive to friction changes of the surface. Particularly, regarding thin sheets like graphene, FFM allows the identification of graphene on different substrates~\cite{Marsden.2013b}, layer number and lattice orientation analysis~\cite{Lee.2010b,Marsden.2013b} as well as the observation of ripple structures in mechanically exfoliated SLG~\cite{Choi.2011}.
\begin{figure*}
	\centering
	\includegraphics[width=\textwidth]{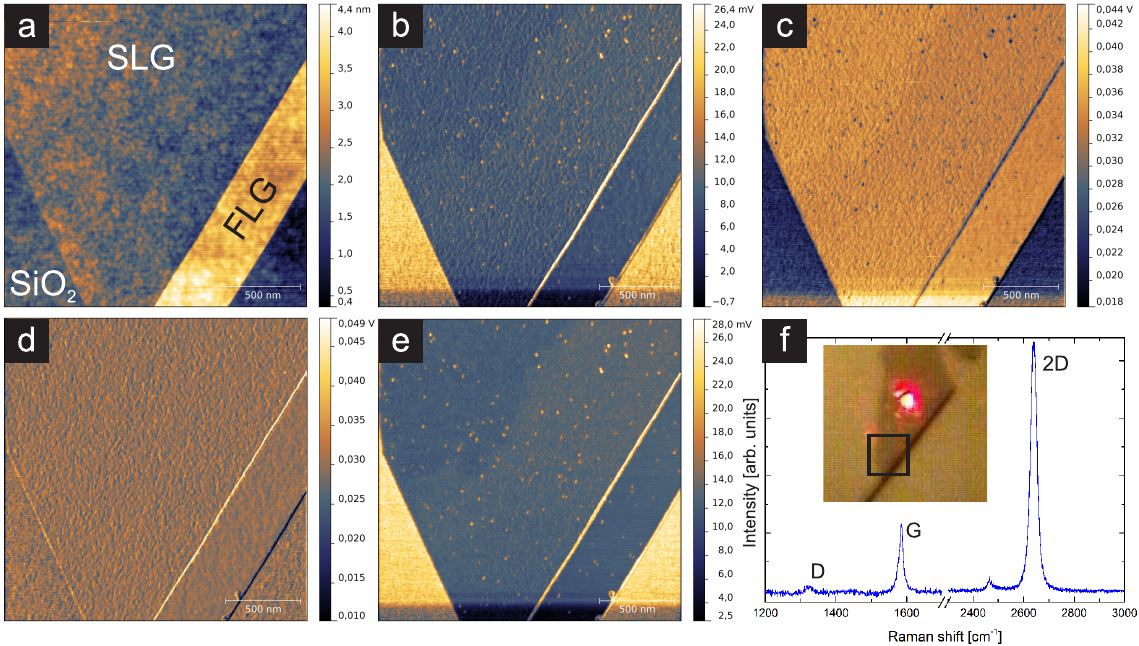}
	\caption{Scanning probe images of a SLG/SiO$_2$ sample irradiated with Xe$^{38+}$ and $E_{\rm kin}$=260~keV. Images were taken with 1 Hz scan speed and typical loading forces of a few nN. (a) topography, (b) friction trace, (c) friction retrace (voltage corresponds to friction force). (d) sum of (b) and (c) yielding topography induced friction. (e) Friction force map. (f) optical microscope image (position of FFM images is marked) and Raman spectrum. The intensity ratio I(2D)/I(G)$\approx$4 and linewidth of G peak FWHM(G)=14.6$\pm$0.3\ cm$^{-1}$ are evidence for SLG. }
	\label{fig:1}
\end{figure*}

After irradiation with varying charge states $q$, topography measurements by conventional contact mode revealed neither in SLG nor in FLG features that could be associated to ion irradiation. Only by using FFM, point-like defects were observed in the FFM friction signal, whose number corresponds very well with the applied ion fluence. We therefore conclude that each defect detected by FFM is related to a single ion impact. As an example we show the result of an ion irradiation experiment with Xe$^{38+}$ on SiO$_2$ supported graphene in figure~\ref{fig:1}. In the topography signal (a), SLG is barely visible due to its low height of nominally 3.95~\AA. However, FFM can easily identify graphene layers by the contrast change in friction trace (b) and retrace (c) signals due to the different friction coefficients $\mu$ for graphene and SiO$_2$, respectively. Because torsion forces of the AFM cantilever lead to a linear deflection of the laser on the position sensitive detector, voltage values correspond qualitatively to friction forces. The sum of signals (b) and (c) yields a topography induced friction signal~\cite{Grafstrom.1994}, here shown in (d). Accordingly, half of the difference of (b) and (c), i.e.~the half width of the trace/retrace loop, represents the friction forces and is called a FFM-map (e) in the following.

\begin{figure}
  \centering
	\includegraphics[width=\columnwidth]{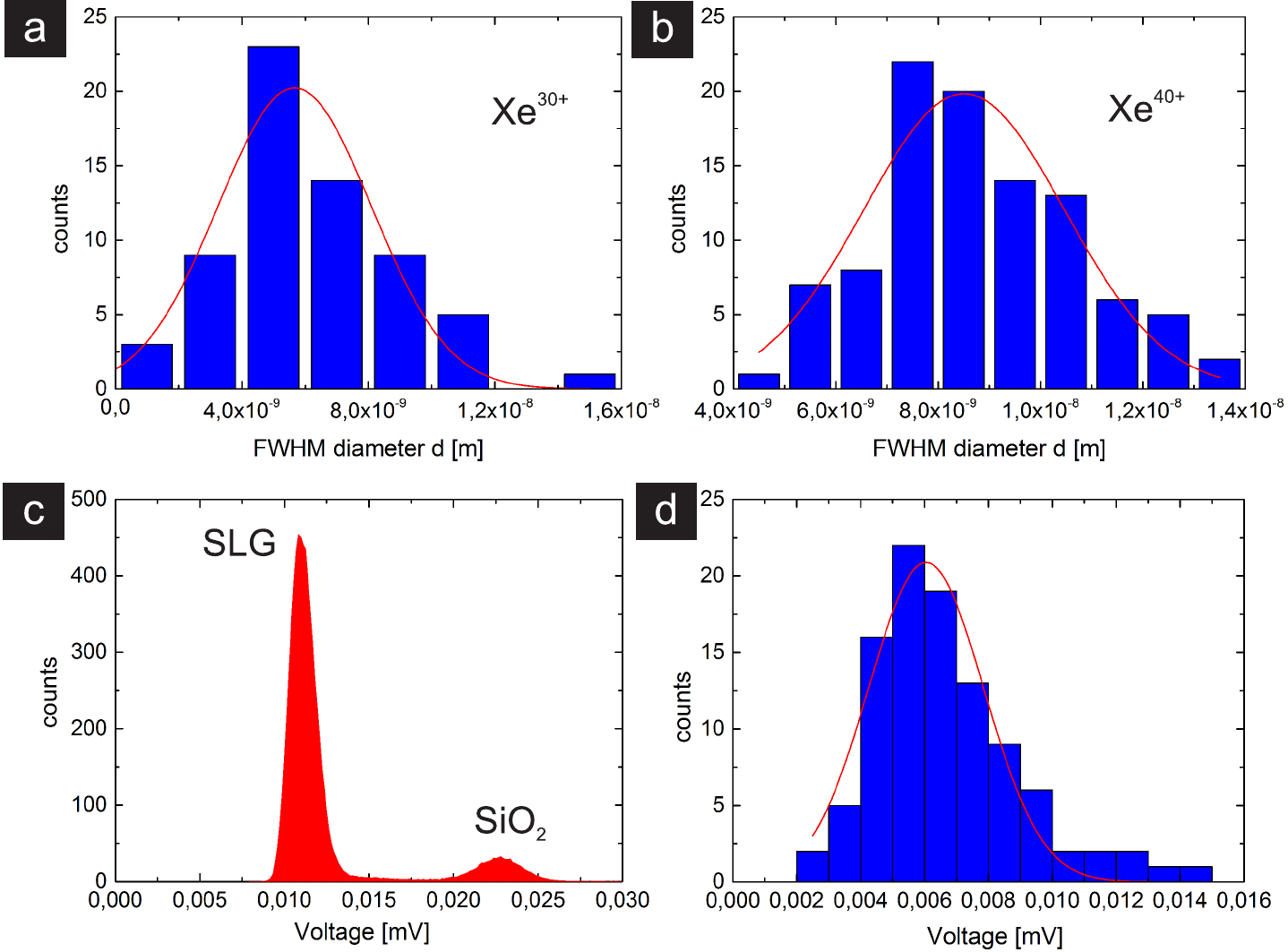}
		\caption{Histograms of defect diameters after irradiation with 260~keV Xe with (a) $q$=30 and (b) $q$=40, respectively. Histograms of friction coefficients for (c) pristine sample and (d) for defective areas (compiled from 100 individual impacts).}
	\label{fig:2}
\end{figure}

By comparing figs.\ref{fig:1}(d) and (e) a clear distinction between different materials, different graphene layers and other features is possible, as they show different friction values corresponding to their respective friction coefficient $\mu$. A detailed analysis reveals a $\mu_{\rm SiO_2}$ with twice the value of SLG (see fig.~\ref{fig:2}c). For few layer graphene, friction forces are somewhat lower than for SLG (e), in accordance with recent work~\cite{Lee.2010b}. The FFM map shows ion induced defects as point-like structures with a clearly enhanced friction compared to friction on SLG. Averaging over 100 defects on SLG yields $\mu_{\rm defects}\approx 1.5\times \mu_{\rm SLG}$ (see fig.~\ref{fig:2}d). The diameter of the ion induced defects was determined by gaussian fits to line scans taken in the FFM images. To rule out tip artefacts, we took with each new tip images from a reference sample (irradiated with $q$=35 at 260~keV). From this image a scaling factor for that tip was determined by which the measured data could be calibrated against the reference. This method ensures that any relative change in diameter will be detected independent on the individual tip shape or size. The mean defect diameter for all irradiated samples was then determined from gaussian fits to the respective diameter histograms. As an example, figure~\ref{fig:2} shows diameter $d$ (FWHM) histograms of HCI induced defects on SLG after irradiation with $q$=30 (a) and $q$=40 (b). 

\begin{figure}
	\centering
	\includegraphics[width=\columnwidth]{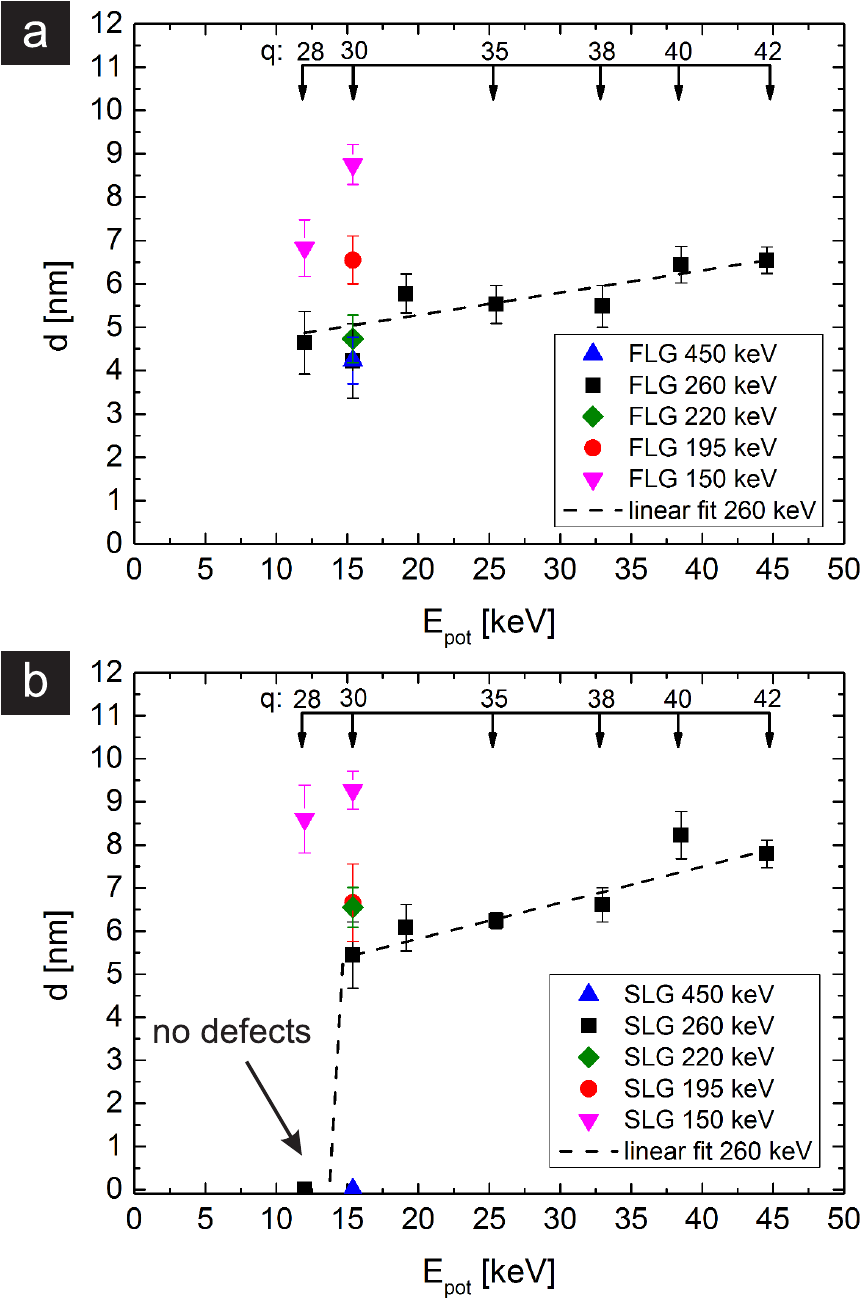}
	\caption{Diameters $d$ of HCI induced defects on FLG (a) and SLG (b), analyzed by FFM for varying charge states $q$ and varying kinetic energy $E_{\rm kin}=(150-450)$~keV) are shown. Defects were found on SLG for $q$=42 down to $q$=30 at E$_{\rm kin}$=260~keV with a threshold between $q$=28 and 30, whereas defects on FLG could be found for every $q$ used here. For decreasing $E_{\rm kin}$ the diameter increases on SLG as well as on FLG. The threshold $E_{\rm pth}$ on SLG shifts to higher $q$ with increasing $E_{\rm kin}$ (see $q$=30, 450~keV) and shifts to lower $q$ with decreasing $E_{\rm kin}$ (see $q$=28, 150~keV).}
	\label{fig:3}
\end{figure}

The results for FLG and SLG for all charge states $q$ used in this study are compiled in figs.~\ref{fig:3}(a) and (b). The error bars represent the FWHM of the gaussian. For both FLG (fig.~\ref{fig:3}a) and SLG (fig.~\ref{fig:3}b), there is a clear correlation between the diameter and the potential energy ($q$), i.e.~a higher charge state $q$ yields an increased diameter $d$. A linear fit of the data for the irradiation series at $E_{\rm kin}$=260~keV reveals a slope of (0.21$\pm$0.04)~nm/keV for SLG and a lower slope of (0.13$\pm$0.03)~nm/keV for FLG. In case of FLG, we could identify HCI induced defects after irradiation with every charge state used here, down to $q$=28. Hence, no threshold behaviour is found for FLG. In contrast, for SLG (see fig.~\ref{fig:3}b), the irradiation series with $E_{\rm kin}$=260~keV (solid squares) clearly shows a threshold behaviour with respect to potential energy. The threshold is between charge states $q$=28 and 30, corresponding to potential energies of 12 keV $<$ $E_{\rm pth}$ $<$ 15.4 keV. That is, a minimum potential energy $E_{\rm pth}$ (or corresponding $q$) is necessary for defect formation in graphene. In addition, the threshold value varies with the kinetic energy as can be seen from fig.~\ref{fig:3}b. At higher kinetic energies $E_{\rm pth}$ shifts to higher values, and to lower values at lower kinetic energies. The correlation between the diameter and the kinetic energy is equally apparent: For FLG (fig.~\ref{fig:3}a) as well as for SLG (fig.~\ref{fig:3}b), we find an increase in defect diameter with decreasing kinetic energy. Note however the distinct deviation between defect diameter in SLG and FLG at higher kinetic energies. 

Let us first briefly discuss the FLG results. Due to our preparation technique via mechanical exfoliation from an HOPG crystal, we assume our FLG samples to be very similar to bulk HOPG samples. Our data should thus be comparable to recent results of Ritter et al~\cite{Ritter.2010,Ritter.2013b}, who reported HCI (Ar$^{9+}$ up to Bi$^{62+}$) induced defects on HOPG. In agreement with our data here, they did not find a threshold, i.e.~defects were created with potential energies as low as $E_{\rm pot}$=0.99~keV. No systemativ influence of the charge state on the diameter of HCI-induced defects measured by AFM was reported. Here, we find a clear change in diameter as a function of the charge state. Intruigingly, the defects become larger when the projectile is slower. This fact could be related either to the time the projectile spends above the individual graphite layer and/or to the number of layers that is affected.

\begin{figure}
	\centering
	\includegraphics[width=\columnwidth]{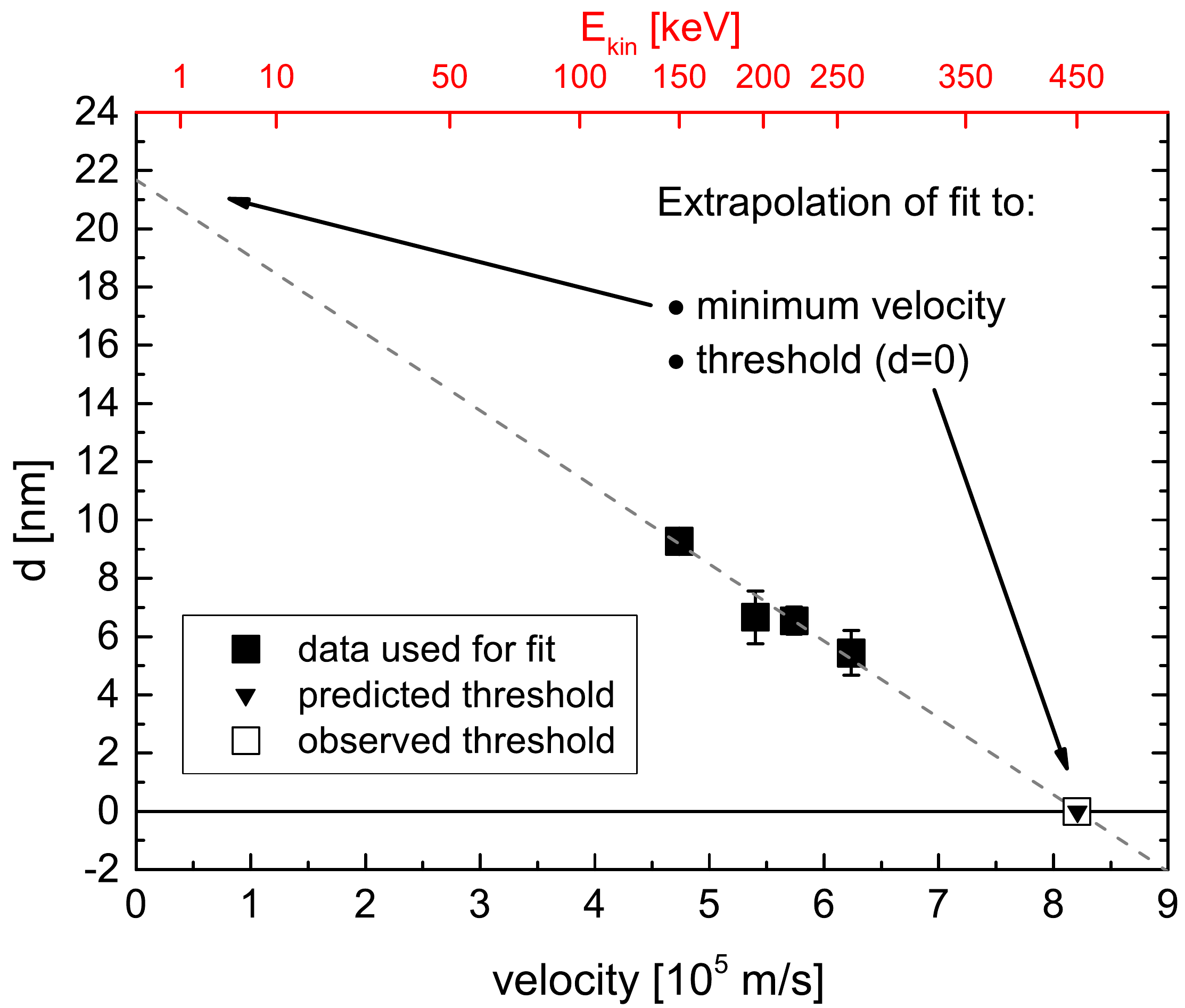}
	\caption{Diameters $d$ of HCI induced defects on SLG at $q$=30 (taken from figure~\ref{fig:3}) as function of the velocity $v$ (for $E_{\rm kin}$ see upper scale). A linear fit to data points for 150~keV $\leq E_{kin}\leq$ 260~keV was performed and extrapolated. The threshold prediction agrees well with our experimental findings.}
	\label{fig:4}
\end{figure}

Next, we want to discuss the results obtained from HCI irradiated SLG. The threshold behaviour seen in fig.~\ref{fig:3}(b) gives clear evidence that the potential energy deposited by the HCI is essential for defect creation. The kinetic energy can however not be completely neglected, as the actual size of the affected area depends strongly on the kinetic energy. This finding is in contrast to what has been observed in bulk alkali halides, alkaline earth halides, and HOPG~\cite{ElSaid.2008,ElSaid.2012b,Heller.2008,Ritter.2013b}. Particularly, for CaF$_2$, the defect (hillocks) size does not depend significantly on the kinetic energy, whereas defect diameters in SLG depend strongly on $E_{\rm kin}$. We therefore propose that in the case of graphene an entirely different mechanism is triggered by the HCI impact. With SLG (in contrast to HOPG) we can safely rule out effects from deeper graphite layers and the SiO$_2$ substrate, which in our measurements did not show any significant defect formation upon HCI irradiation. We can thus directly conclude that for SLG the key factor for defect size must be the time, during which the projectile interacts with the graphene layer. A velocity plot reveals a linear dependency of $d$ on the velocity $v$ of the incoming Xe$^{30+}$ion (see fig.~\ref{fig:4}). The linear fit to the data points of observed defects at lower kinetic energies (150--260~keV) can be extrapolated to higher velocities and predicts that no defects should be created at $E_{\rm kin}$=450~keV, which is in perfect agreement with our measured data. Because defect diameters in SLG and FLG increase with increasing $q$ at fixed $E_{\rm kin}$, the defect formation must be driven by the potential energy deposition into the surface. From the observed decrease of $E_{\rm pth}$ with decreasing $E_{\rm kin}$ (see fig.~\ref{fig:3}b, $q$=28 and $q$=30 on SLG), we conclude that the available potential energy is deposited more efficiently by slower HCI. To fully describe the interaction of HCI with graphene we therefore have to take into account both forms of energy deposition. For simplicity we neglect defect formation by nuclear collisions in the following as this has been shown to create mainly single and double vacancies~\cite{Lehtinen.2010}. 

In general, the HCI-solid interaction process can be separated into several stages \cite{Arnau.1997,Schenkel.1999b}. The first stage is usually explained in terms of the over-the-barrier model~\cite{Burgdorfer.1991}. As the ion approaches the surface and passes a critical distance $r^{\rm crit}=\sqrt{2q}/W_{\phi}$ depending on the work function $W_{\phi}$ of the respective material~\cite{Winter.1996}, it starts to neutralize by resonant electron capture and Auger ionisation. For single layer graphene on SiO$_2$, the work function is typically 4.8~eV and this distance would thus yield $r^{\rm crit}\simeq (2-3)$~nm~for the charge states investigated here. During the neutralization phase, electrons from the target get caught into highly excited states of the projectile resulting in a so called hollow atom~\cite{Briand.1990}. Deexcitation of this hollow atom starts in front of the surface and may continue in graphene via different Auger processes as well as collisional electron-electron processes. As a consequence, many excited electrons in the low energy range up to a few hundred eV as well as keV electrons due to Auger decay of unbalanced holes are created. 
The potential energy of each HCI results thus in a strong excitation of the electronic system as well as significant depletion of electrons of graphene within a small area of only a few nm in diameter. This phase is directly related to the kinetic energy via the time of flight. The slower the ion, the more time it spends in this region resulting in an increased number of electrons emitted from graphene, provided the excitation processes take place on a time scale that is comparable with the time of interaction. Typical time scales for the above-surface processes are on the order of a few femtoseconds~\cite{Winecki.1996,Burgdorfer.1993} which is the same time a projectile with a few 100 keV in kinetic energy needs to overcome a distance of a nanometer. We therefore propose that the defect creation is governed mainly in the flight phase, within which the ion is neutralized by electrons supplied by graphene. %In order to differentiate further between the respective contributions from above- and below-graphene one would need to go to extremely low kinetic energies which is currently not possible with our set-up.

Finally, we address the nature of the defects. Our data clearly shows that permanently modified regions can be created in graphene by HCI irradiation. However, the nanoscaled size of the modified areas prevents an easy identification. In principle there are several mechanisms by which these final modifications could be created. The energy from the electronic system may be transferred via electron-phonon coupling to the lattice, resulting in a thermal spike \cite{Toulemonde.1992} and possibly melting of the target~\cite{Aumayr.2011}. This scenario works perfectly well for defect formation in various bulk targets~\cite{ElSaid.2012b,Heller.2008,Aumayr.2011,Lake.2011b}. For graphene this scenario is questionable, as any form of electronic excitation should dissipate rapidly due to the high carrier mobility, and secondly, an inefficient electron-phonon-coupling on the order of $10^{13}~\rm{J/(K\cdot s\cdot m)}$~\cite{Fong.2012}) combined with the unusually high thermal conductivity of graphene should prevent any structural modifications related to heating. Alternatively, the strong perturbation of the electronic system may result in direct removal of atoms via Coulomb repulsion or non-thermal melting~\cite{Aumayr.2011}. To our knowledge there has been no clear evidence yet that either one of the latter two mechanisms can be triggered by HCI. A more likely mechanism is that the strong depletion of electrons in a localized area of SLG leads to a weakening or breaking of chemical bonds. The resulting dangling carbon bonds act as attractive adsorption sites for adatoms such as oxygen and hydrogen. This again will yield changes such as charge redistribution and rehybridization. Such chemically modified graphene will then give rise to an enhanced friction signal ~\cite{Ko.2013,Byun.2011}. Note, that the absolute value of friction enhancement depends on the exact experimental conditions. A quantitative comparison with our data is thus difficult, but enhanced friction without topography enhancement clearly points towards hydrogenated graphene~\cite{Byun.2011}. The size of such a modified area would be directly related to the number of depleted electrons and, hence, for a given charge state to the velocity of the incoming ions. Further corroboration of this model is given by electron emission data on HCI impinging on HOPG~\cite{Wang.2011}. Electron emission yields $\gamma$ were taken as function of $E_{\rm pot}$ for different $E_{\rm kin}$. As a crucial result, $\gamma$ increases linearly with increasing $E_{\rm pot}$ and $\gamma$ decreases with increasing velocity, which is in excellent agreement with our data for $d(v)$, see fig.~\ref{fig:4}.

To conclude, we showed that nanosized defective areas in single layer graphene are induced by individual highly charged ions. The size of the defective area depends on the potential energy of the HCI and, surprisingly, on the kinetic energy as well. We propose that the significant depletion of electrons during HCI approach yields a breaking of graphene bonds, most likely followed by a hydrogenation. As a consequence the time of flight above the surface is the key parameter, by which defect size and threshold value are set. Defect creation in graphene by HCI can thus be controlled by a simple variation of kinetic energy (for a given and sufficiently high charge state). According to the extrapolation in fig.~\ref{fig:4}, rather large defects with diameters of up to 20 nm could be created by using very slow HCI ($E_{\rm kin}<$ 1 keV). This would also offer the chance to investigate the exact nature of the defects as well as their potential for applications in much more detail. 

This work has been financially supported by the DFG within the SFB 616: {\it Energy dissipation at surfaces} (J.H.) and SPP 1459: {\it Graphene} (R.K.) as well as by the European Community as an Integrating Activity 'Support of Public and Industrial Research Using Ion Beam Technology (SPIRIT)' under EC contract no. 227012.

\bibliographystyle{unsrt}
\bibliography{paper}

\end{document}